\documentclass[11pt]{article}

\usepackage{epsf}
\usepackage{graphicx}
\usepackage{amssymb}
\usepackage{setspace}

\begin{document}
\newcommand{\be}{\begin{equation}}
\newcommand{\ee}{\end{equation}}
\newcommand{\bea}{\begin{eqnarray}}
\newcommand{\eea}{\end{eqnarray}}

\title{Interstitial fractionalization and spherical crystallography}

\author{\small \\ Mark J. Bowick$^{1}$\thanks{\tt
bowick@physics.syr.edu} \, David~R. Nelson$^{2}$\thanks{\tt
nelson@cmt.harvard.edu} \, and Homin Shin $^{1}$\thanks{\tt
hshin@physics.syr.edu}
\\ $^1$Physics Department, Syracuse University,\\
Syracuse, NY 13244-1130, USA
\\ $^2$Lyman Laboratory of Physics, Harvard University, \\
Cambridge, MA 02138, USA \\ }

\maketitle

\begin{abstract}
Finding the ground states of identical particles packed on spheres
has relevance for stabilizing emulsions and a venerable history in
the literature of theoretical physics and mathematics.  Theory and
experiment have confirmed that defects such as disclinations and
dislocations are an intrinsic part of the ground state.  Here we
discuss the remarkable behavior of vacancies and interstitials in
spherical crystals.  The strain fields of isolated disclinations
forced in by the spherical topology literally rip interstitials and
vacancies apart, typically into dislocation fragments that combine
with the disclinations to create small grain boundary scars.  The
fractionation is often into three charge-neutral dislocations,
although dislocation pairs can be created as well.  We use a
powerful, freely available computer program to explore interstitial
fractionalization in some detail, for a variety of power law pair
potentials.  We investigate the dependence on initial conditions and
the final state energies, and compare the position dependence of
interstitial energies with the predictions of continuum elastic
theory on the sphere.  The theory predicts that, before
fragmentation, interstitials are repelled from 5-fold disclinations
and vacancies are attracted.  We also use vacancies and
interstitials to study low energy states in the vicinity of ``magic
numbers" that accommodate regular icosadeltahedral tessellations.
\end{abstract}

\section{Introduction}
Producing stable emulsions of two immiscible fluids, such as oil and
water, is a challenging and important problem, not only because of
technical applications but also from the perspective of fundamental
science. One strategy for stabilizing emulsions, dating back at
least 100 years~\cite{Pick}, involves coating droplets of one phase
with small colloidal particles to impede droplet
coalescence~\cite{Sac}. The colloidal ``armor plating" of these
Pickering emulsions also plays a role in colloidosomes,
colloid-coated lipid bilayer vesicles used for encapsulation and
delivery of flavors, fragrances and drugs~\cite{Dinsmore}. Identical
micron-sized particles tend to crystallize under typical
experimental conditions, and it is of some interest to understand
the defect structure of locally crystalline ground states on a
sphere, since these influence the strength of the colloidal armor.
Understanding particle packings on a sphere also involves
fundamental questions of theoretical physics and mathematics, dating
back to work by J. J. Thomson in 1904~\cite{Thoms}. Although Thomson
was interested in electrons interacting with a repulsive $1/r$
potential~\cite{Dogson,Dog2,Toomre, Moore}, the problem of
determining crystalline ground states on a sphere can be posed more
generally in terms of continuum elastic theory, Young's moduli and
defect core energies, for particles interacting with a wide variety
of pair potentials~\cite{BNT, BCNT2002,BCNT2006}.

Defects play an essential role in describing crystalline particle
packings on the sphere. At least twelve particles with 5-fold
coordination (i.e., 12 disclinations) are required for topological
reasons, and like the 5-fold rings in carbon fullerenes, one might
expect that the energy would be minimized if the disclination
positions approximated the vertices of a regular icosahedron. This
expectation, which also plays a role in geodesic domes and in the
protein capsomere configurations of spherical virus
shells~\cite{Caspar,zandi}, is nevertheless \emph{violated} when the
shells are sufficiently large and disclination buckling~\cite{LMN}
out of the spherical environment is suppressed by surface tension.
Consistent with theoretical expectations~\cite{Toomre,Moore,BNT},
experiments on particle-coated water droplets in oil~\cite{science}
reveal that the twelve excess disclinations sprout grain boundary
``scars" for sufficiently large $R/a$, where $R$ is the sphere
radius and $a$ is the average particle spacing. When triangulations
of microscopic particle packings are used to reveal the local
coordination number, these grain boundaries appear as additional
dislocations, i.e., 5-7 pairs, arrayed around an unpaired 5, in a
pattern such as 5-7--5-7--5--7-5--7-5. Although the critical value
of  $R/a$ above which grain boundaries appear depends on microscopic
details, both theoretical estimates~\cite{BNT} and
experiments~\cite{science} indicate that this instability arises as
soon as $R/a \gtrsim 5-6$, i.e., when the total number of particles
exceeds several hundred. Thus, unlike crystals in flat space,
dislocations arrayed in grain boundaries are an intrinsic part of
the ground state.   These grain boundaries can, moreover, stop and
start freely on the sphere, unlike their flat space counterparts.
Such terminations occur naturally (and with low energy cost) because
crystalline grains rotate under parallel transport due to the
nonzero Gaussian curvature of the sphere.

If disclinations and dislocations are crucial for understanding
spherical crystallography, what can we say about \emph{vacancies and
interstitials}, which are well known to play a key role in
conventional crystals~\cite{Mermin}?    It is natural to introduce
vacancies and interstitials in an attempt to understand spherical
particle packings that deviate from certain ``magic numbers"
$N_{nm}$. These preferred particle numbers, introduced by Caspar and
Klug in their analysis of viral shells~\cite{Caspar}, refer to
special commensurate spherical tessellations, indexed by a pair of
integers $(n,m)$. The corresponding particle number associated with
these commensurate particle packings is $N_{nm}=10(n^2+m^2+nm)+2$.
It is tempting to ignore the instability to grain boundary scars for
large $N_{nm}$ and regard the commensurate $(n,m)$ lattice as an
interesting metastable state. It would then be natural to introduce
vacancies and interstitials to describe candidate ground state
packings for $N_{nm} \pm t$ particles, where $t=1, 2,...$ and much
less than the distance to the next magic number.   There is,
however, another surprise in store: In contrast to flat space, where
vacancy and interstitial defects are stable and well defined, we
find that interstitials and vacancies are typically ripped apart
into dislocation fragments by the strain fields of nearby 5-fold
disclinations. The dislocations then combine with some of the excess
5's to form defect clusters such 7-5-7. Thus, vacancies and
interstitials lose their integrity via fragmentation in spherical
crystals, and mediate formation of small grain boundary scars.    A
full account of the ground states on the sphere for $N_{nm} \pm t$
particles is beyond the scope of this paper.   We present here,
however, a study of the fragmentation process itself.  We
concentrate on interstitials for simplicity. Although vacancies
behave in a similar fashion and can be studied by the same computer
program~\cite{Java}, the theoretical analysis is more complicated
for vacancies, which are typically crushed by elastic forces into
objects with a low two-fold symmetry, even in flat
space~\cite{Cockayne, Jain}. In addition, we compare our results for
interstitial energies (which are position-dependent) to predictions
of continuum elastic theory and provide information about the
energetics in the vicinity of $(n,m)$ commensurate spherical
tessellations.

We begin by studying in Section~\ref{2} the fractionalization of
interstitials in two-dimensional curved crystals via numerical
simulations of the generalized Thomson problem~\cite{Java}.
Interstitials (or vacancies) in a sufficiently large spherical
crystal are, unlike in flat space, unstable to unbinding into
several individual dislocations, each of which glides towards the
nearest disclination, eventually forming the small grain boundary
scars mentioned above~\cite{BST}. In this section, we investigate
such unstable interstitials and their fractionalization in some
detail. We determine the energy of interstitials before and after
fragmentation as a function of their position relative to the twelve
excess disclinations that are inevitably present in a spherical
crystal in Sections~\ref{3} and ~\ref{4}. In Section~\ref{5}, we
calculate the interaction energy of interstitials with the extra
disclinations within continuum elasticity theory. Finally, we
discuss ground state energies close to preferred ``magic numbers" of
particles on the sphere corresponding to the regular icosadeltahedra
of Caspar and Klug~\cite{Caspar}.

\begin{figure}[!h]
\centering
\includegraphics[scale=0.6]{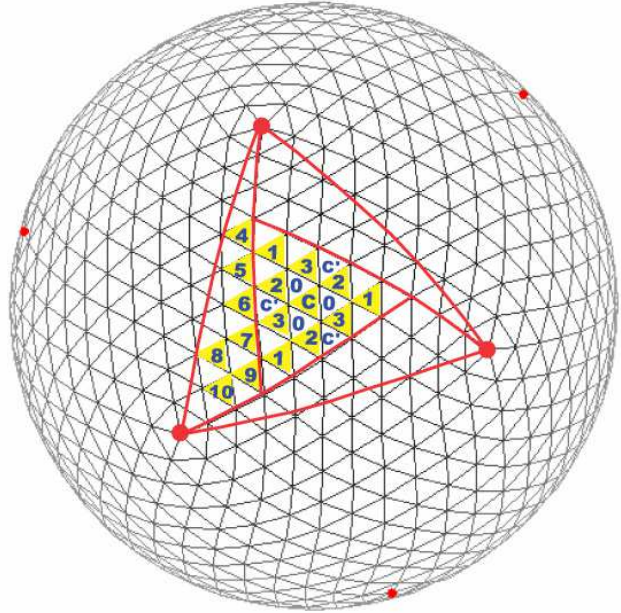}
\caption{Initial icosahedral configuration of an (8,3) spherical
crystal lattice with 12 disclination defects.  This lattice is
chiral, and is arranged such that 8 steps along a Bragg row,
followed by 3 steps to the right along a Bragg row at a
$120^{\circ}$ angle, connect neighboring 5-fold disclinations.
Distinct initial locations for interstitials are shown as triangular
plaquettes labeled by characters and numbers.} \label{initial}
\end{figure}

\section{Interstitial fractionalization}
\label{2}

An icosahedral spherical crystalline lattice (a regular
icosadeltahedron) can be constructed for every pair of integers
$(n,m)$, where the number of vertices or particles in the
tessellation is given by the ``magic numbers" $N_{nm}$: \be
\label{PQ} N_{nm}=10(n^2 + nm + m^2) + 2 \ . \ee  We are interested
in studying the effect of inserting an interstitial or a vacancy
into a regular $(n,m)$ icosahedral lattice by adding or removing a
single particle, giving rise to particle numbers $N_{nm}\pm1$ not
falling in the classification of Eq.~(\ref{PQ}) and in studying the
relation of interstitials to the extra dislocation defects (scars)
found in spherical crystals above a critical system size
\cite{BNT,science}. Refs.~\cite{Cockayne, Jain, Fisher, Frey, Perts,
Libal} study configurations and energies of interstitial and vacancy
defects and their energetics in triangular lattices in flat space.

\begin{figure}[t]
\centering
\includegraphics[scale=0.5]{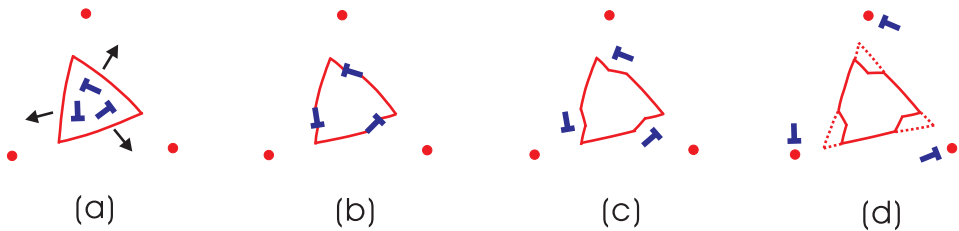}
\caption{A schematic of interstitial fractionization. The $\dashv$
symbols are alternative ways of representing dislocations depicted
elsewhere as 5-7 pairs.} \label{fracschematic}
\end{figure}

\begin{figure}[!h]
\centering
\includegraphics[scale=0.7]{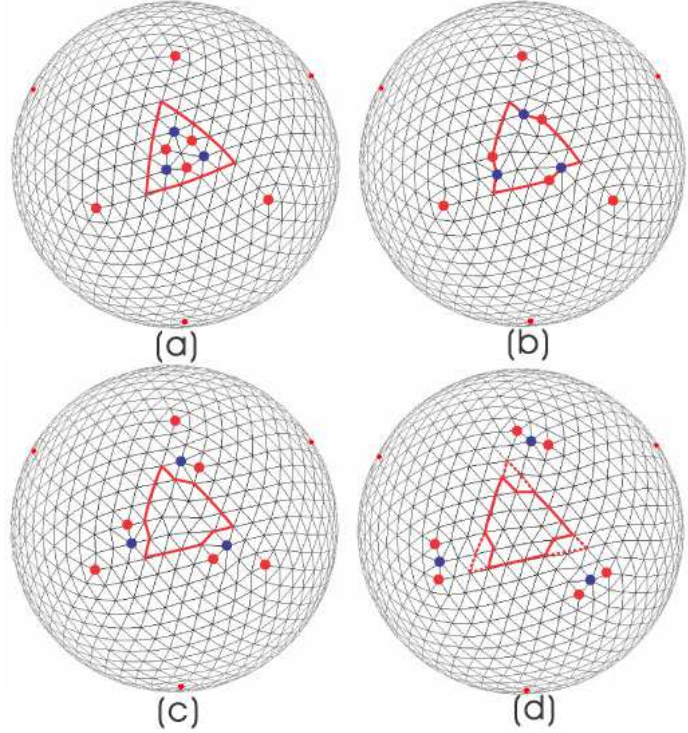}
\caption{The fractionalization of an interstitial defect inserted at
the center of three neighboring disclinations in an (8,3) spherical
tessellation: (a) the initial interstitial defect configuration
($I_3$) shown surrounded by a triangular reference contour; (b) the
bound defect unbinds to form three separate dislocations (5-7)-
pairs; (c) each dislocation glides (i.e., moves parallel to its
Burgers vector) towards the nearest disclination; (d) three
mini-grain boundary scars are formed. We also keep track of the
evolution of the initial defect in (a) by illustrating the
deformation of the triangular contour around the initial defect for
(b), (c), and (d) induced by the passage of the dislocation.}
\label{frac}
\end{figure}

The presence of excess dislocation defects in the ground state of
spherical crystals is dramatically illustrated by the following
numerical experiment: we start with a regular icosadeltahedral
tessellation of the sphere {--} say an $(8,3)$, corresponding to
$N_{83}=972$ (Fig.~\ref{initial}). This may be done with the applet
located at \cite{Java} using the Construct $(m,n)$ algorithm.
Although the true ground state for $972$ particles on the sphere
with most pair potentials has additional dislocation defects (i.e.
tightly bound pairs of 5- and 7- coordinated particles) arrayed in
grain boundary scars~\cite{BNT}, the regular icosadeltahedral
lattice is a local minimum from which it is difficult to escape
without the addition of thermal noise. In fact it is a major
challenge to find fast and reliable algorithms to locate the true
ground state (global minimum) in this problem with its complex
energy landscape. Now add a single particle to the lattice at the
center of mass of a spherical triangle whose vertices are 3
nearest-neighbor 5-fold disclinations (shift + click). The {\em
self-interstitial} so formed is then relaxed by a standard
relaxation algorithm, with sufficient thermal noise to allow
dislocation glide over the Peierls potential \cite{NatMat}. One
immediately finds that an interstitial is structurally unstable. In
a few time steps it morphs into a complex of dislocations with zero
net Burgers vector. The most common structure observed is a set of
three dislocations, with Burgers vectors perpendicular to a line
joining each 5-7 pair, inclined at $120^{\circ}$ angles to each
other. Eventually the interstitial complex is ripped apart entirely,
as illustrated schematically in Fig.~\ref{fracschematic} (see also
Fig.~\ref{frac}). Intermediate configurations and final states as
the dislocations glide apart will be classified later. Most often
three separate dislocations are formed which each glide toward a
$5$-fold disclination. The end result is the formation of a
``mini-scar" (a $5-7-5$ grain boundary) at each of the vertex $5$s.
Subsequent removal of a particle to restore the particle number to
the original $972$ and relaxing still leaves scars with total energy
lower than the starting configuration with $12$ isolated $5$'s. This
observation confirms that scars are definitely lower energy states
and not simply artifacts of the relaxation algorithm.

The above phenomenon of low-temperature ($T \gtrsim 0$) {\em
unbinding} of dislocations by spatial curvature is a curved space
analog of melting at finite temperature. The extended nature of
fractionated interstitials (each separating dislocation component
involves an extra row of particles) means that they cannot be
treated as small perturbations from the initial spherical crystal
with particle number $N_{nm}$.

Let's return to the specific case of the $N=973$ particle
configuration generated by an interstitial inserted into one
triangular plaquette of a regular $(8,3)$ lattice of $N_{nm}=972$
particles with the requisite $12$ disclination defects (5-fold
coordinated particles) at the vertices of a regular icosahedron, as
shown in Fig.~\ref{initial}. The spherical crystal is distorted by
the additional particle {--} the local configuration adopted by the
interstitial changes as the crystal relaxes toward a lower energy
state. As in the case of planar lattices, we also find here that the
various interstitial defect configurations appear, such as the
twofold symmetric interstitial $I_2$, the threefold symmetric
interstitial $I_3$, and the fourfold symmetric interstitial $I_4$
(see Figs.~\ref{frac},~\ref{rot2}, and ~\ref{rot_I_4_2}). The most
common intermediate complex formed by the interstitial is threefold
symmetric in the rough form of a triangular loop composed of three
dislocations with radially oriented Burgers vectors. All of the
configurations adopted by an interstitial prior to unbinding can be
described as a set of dislocations with zero net Burgers vector.

In marked contrast to interstitial defects in a planar crystal,
interstitial defect configurations in a spherical crystal are
metastable states with characteristic decay processes. As we shall
see, the instability is caused by interactions with the inevitable
disclinations associated with the nonzero Gaussian curvature of the
sphere. A representative evolution of an interstitial inserted at
the center of three neighboring disclinations in an $(8,3)$
icosadeltahedron is shown in Fig.~\ref{frac}. After some local
relaxation the interstitial configuration denoted $I_3$ is formed,
as shown in Fig.~\ref{frac} (a). We also show there the construction
of a triangular contour surrounding the original interstitial
defect. The presence of an interstitial follows because the contour
encloses $7$ particles instead of $6$, the number appropriate to a
perfect triangular lattice. During the annealed relaxation
illustrated in Figs.~\ref{frac}(b) through (d), the dislocations
which are bound together in the initial interstitial unbind into
individual dislocations and subsequently glide towards nearby
disclinations, eventually forming minimal 5-7-5 grain boundary
scars. If one thinks of the initial dislocations as internal degrees
of freedom within the interstitial, one could say that one-third of
an interstitial is present in each mini-scar in the final state and
thus that the interstitial demonstrates $1/3$ fractionalization. For
other initial conditions the fractionation is into two dislocations,
each representing $1/2$ of the original interstitial. The
instability of interstitial defects in curved crystals may be
studied via continuum elasticity theory by calculating the
interaction energies between defects at each stage of the relaxation
process of Fig.~\ref{frac}~\cite{BST}. We note that the triangular
plaquette around the initial defect has been deformed such that it
conforms as closely as possible to the regular triangular lattice
during the relaxation process. Its deformation reveals the passage
of the escaping dislocations.

\section{Interstitial defect energies}
\label{3}

In this section, we discuss the energy of an interstitial defect in
a spherical crystal. Consider $N$ point particles constrained to lie
on the two-dimensional surface of a unit sphere. The energy of $N$
particles interacting through a generalized Coulomb potential within
the curved surface is given by \be
\label{general}E=\frac{1}{2}\sum_{i,j} \frac{1}{|{\bf r}_i-{\bf
r}_j|^s} \ , \ee where ${\bf r}_i$ is the position of the particle
in three dimensions and $s$ is an integer. For a flat triangular
lattice with periodic boundary conditions, the interstitial defect
energy at constant density was defined in~\cite{Fisher, Frey} as \be
E_I = E_{relaxed} - E_{per} \ , \ee where $E_{relaxed}$ is the
relaxed energy of the rearranged lattice of $N$ particles with the
interstitial defect in the area $A$, and $E_{per}$ is the energy of
the perfect crystal at the same areal density $N/A$. In curved
space, however, the definition of a ``perfect crystal" is more
subtle, since disclination defects resulting from the Gaussian
curvature and the topology are inevitable. We will take as a
reference crystal the $(n,m)$ icosadeltahedral configurations
corresponding to triangular tessellations of a magic number of
particles $N_{nm}= 10 (n^2 + m^2 + nm) + 2$. Once an interstitial or
vacancy is added to such a $(n,m)$ configuration, we are no longer
at a magic number of particles since these are quite sparsely
distributed. We thus need to define the energy of the perfect
crystal.

Here we define the energy of the interstitial (vacancy) defect at
constant density in the spherical crystal as \be \label{E_I} E_I =
E_{local} - E^*_{annealed} \ , \ee where $E_{local}$ measures the
energy of the relaxed interstitial while the constituent
dislocations are still bound and $E^*_{annealed}$ is the minimum
energy of all possible final states attained after \emph{annealed
relaxation} leading to interstitial {\em fractionalization}. This
definition will be more explicitly discussed in the following
section (see Table~\ref{anneal}). We note that both $E_{local}$ and
$E^*_{annealed}$ are measured at the same areal density $(N_{nm} \pm
1)/A$, where $\pm 1$ correspond to an interstitial (vacancy)
respectively. The lowest relaxed energy $E^*_{annealed}$ plays the
role of the energy of the perfect lattice in the planar case at the
density of $(N_{nm} \pm 1)/A$.

We have performed numerical measurements of $E_{local}$ and
$E^*_{annealed}$ for the power-law potentials with $s = 1$, $3$, $6$
and $12$, by adding one interstitial at the center of a spherical
triangle formed by three nearest-neighbor disclinations in the
$(8,3)$ lattice (the location represented by $C$ in
Fig.~\ref{initial}). $E_{local}$ is measured by quenching the system
at the moment just prior to the fractionation of the interstitial
into individual dislocations ((a) in Fig.~\ref{frac}). The results
are reported in Table~\ref{s}.

\begin{table}[!h]
\centering
\begin{tabular}{c c c c c}
\hline\hline
s & 1 & 3 & 6 & 12 \\
\hline
local    & 456601.99 & 2840600.7 & 9.62182 $\times 10^8$ & 3.03015 $\times 10^{14}$ \\
annealed & 456600.91 & 2840025.5 & 9.60570 $\times 10^8$ & 3.00313 $\times 10^{14}$ \\
$E_{local}-E^*_{annealed}$ & 1.08 & 575.2 & 0.01612 $\times 10^8$ &
0.02702 $\times 10^{14}$ \\\hline\hline
\end{tabular}
\caption{The lowest local and annealed relaxed energy with the
central interstitials created by putting a particle at $C$ in
Fig.~\ref{initial} of the $(8,3)$ lattice, for $s=1$, $3$, $6$, and
$12$. The differences between two relaxed energies are calculated.
Because the particles are embedded in a sphere of unit radius with
our conventions, near-neighbor particle spacings are of order $a
\sim N^{-1/2} \ll 1$, leading to a strong $s$ dependence in the
total energy given by Eq.~(\ref{general}).} \label{s}
\end{table}

\section{Position dependence of interstitial defect energies}
\label{4}

By adding a particle in different plaquettes within the large
spherical triangle of Fig.~\ref{initial}, one can explore the
dependence of the final state on the initial interstitial location.
In contrast to the case for planar crystals, both the {\em location}
and {\em orientation} of the interstitial defect relative to nearby
disclinations influences the resultant configuration and its
corresponding evolution, leading to distinct relaxed configurations.

The insertion of a particle at the center of the large spherical
triangle leads to an $I_3$-type initial configuration, whereas
adding the interstitial to the plaquette along an edge results in an
$I_2$-type initial configuration.

During the relaxation process, we also find that the dislocation
complex representing an interstitial can rotate so as to reorient
the Burgers vectors so that constituent dislocations can glide
towards a nearby disclination and bind to it. This phenomenon is
especially noticeable if we place one extra particle slightly off
from the absolute center $C$, such as the locations $C'$ or $0$ in
Fig.~\ref{initial}.

\begin{figure}[b!]
\centering
\includegraphics[scale=0.5]{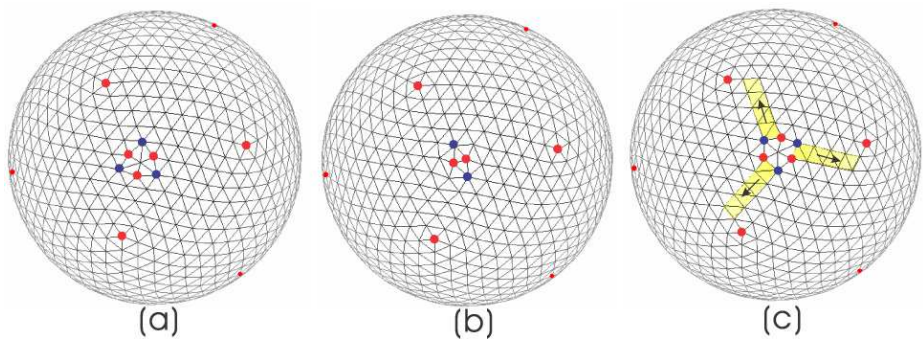}
\caption{The rotational motion of an interstitial configuration
(created with the initial location $C'$ in Fig.~\ref{initial})
mediated by the transition: $I_3 \rightarrow I_2 \rightarrow I_3$}
\label{rot2}
\end{figure}

\begin{figure}[t]
\centering
\includegraphics[scale=0.5]{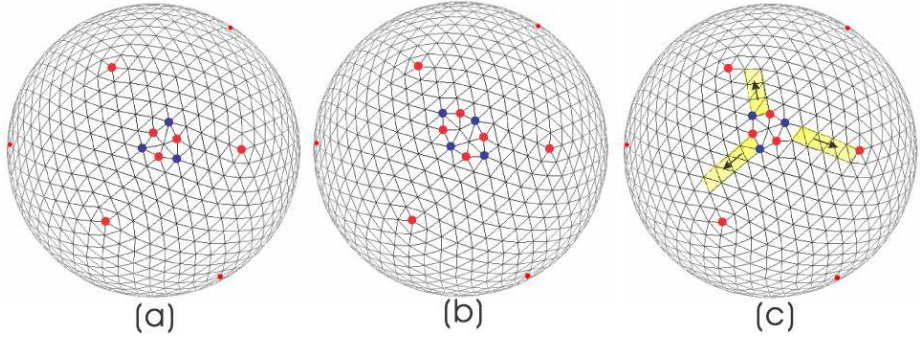}
\caption{The rotational motion of an interstitial configuration
(created with the initial location $0$ in Fig.~\ref{initial})
mediated by the transition: $I_3 \rightarrow I_4 \rightarrow I_3$}
\label{rot_I_4_2}
\end{figure}

In Fig.~\ref{rot2} and Fig.~\ref{rot_I_4_2} we illustrate this
phenomenon more explicitly. In Fig.~\ref{rot2}(a), an interstitial
initially placed at the position $C'$ in Fig.~\ref{initial} morphs
quickly to an $I_3$ configuration. In (a), however, the orientations
of the dislocations within $I_3$ are not appropriate for dislocation
glide to the surrounding disclinations since the $5$ end of the
dislocations point towards rather than away from these 5-fold
disclinations, causing them to be repelled. Glide-induced
fractionalization is therefore prohibited in this orientation.
Remarkably, though, the entire complex of dislocations can change
its orientation by a transition through an intermediate $I_2$
configuration (shown in Fig.~\ref{rot2}(b)) and subsequently to a
second $I_3$ configuration (Fig.~\ref{rot2}(c)). The final $I_3$
configuration is rotated by $60^{\circ}$ with respect to the first
$I_3$ and can now fractionate analogously to an interstitial with
initial position $C$ in Fig.~\ref{initial}.

We also find rotational reorientation of an interstitial defect with
an $I_4$-type intermediate state, as shown in Fig.~\ref{rot_I_4_2}.
This particular relaxation process reveals an interesting feature of
dislocation dynamics on a curved surface. The $I_3$ configuration
generated after the intermediate (see Fig.~\ref{rot_I_4_2}(c)) now
has one dislocation with its glide plane such that it can  glide
head-on into a vertex disclination. This disclination absorbs the
dislocation but hops over one lattice spacing to accommodate the
curved space Burgers vector. The other two dislocations end up bound
in the form of mini-scars. In this case then the interstitial has
fractionated into 2 rather than 3 parts and one say that there has
been $1/2$ fractionization of the interstitial. Absorption and
emission of dislocations by 5-fold disclinations are somewhat
analogous to absorption and emission of vacancies and interstitials
by dislocations (allowing dislocations to climb), a phenomenon
well-known in flat space.

We next catalog the dependence of the final state on the initial
location of the interstitial.  The distinct initial conditions shown
in Fig.~\ref{initial} lead to three different final annealed states,
as summarized in Table~\ref{anneal}. We have checked that all other
possible initial conditions, not indexed in Fig.~\ref{initial}, also
produce one of the listed final states. The final state with three
mini-scars of the form 5-7-5 has the lowest energy of all final
states and provides a measure of $E^*_{annealed}$.

\begin{table}[!h]
\centering
\begin{tabular}{c c c}
\hline\hline
initial location & $E_{annealed}$ & annealed state \\
\hline
C, C', 7, 8, 10 & 9.60570 $\times 10^8$ & 3 scars (5-7-5) \\
0, 2, 3, 4, 6, 9  & 9.61011 $\times 10^8$ & 2 scars (5-7-5) \\
1, 5 &  9.61062 $\times 10^8$ & 2 scars (5-7-5-7-5) \\
\hline\hline
\end{tabular}
\caption{The three classes of final annealed state depending on the
initial interstitial location. The relaxed energies are measured for
the power law potential $s = 6$. Here $E_{annealed}$ with 3 scars
corresponds to $E^*_{annealed}$.} \label{anneal}
\end{table}

\begin{table}[b!]
\centering
\begin{tabular}{c c c c}
\hline\hline
n & transition & $E_{local}$ of $I_3$ & $E_{local}-E^{*}_{annealed}$ \\
\hline
C & $I_3$ & 9.62182 $\times 10^8$ &0.01612 $\times 10^8$ \\
C'& $I_3 \rightarrow I_2 \rightarrow I_3$ & 9.62195 $\times 10^8$ & 0.01655 $\times 10^8$\\
0 & $I_3 \rightarrow I_4 \rightarrow I_3$ & 9.62274 $\times 10^8$ & 0.01708 $\times 10^8$\\
1 & $I_3 \rightarrow I_2$ &  9.62391 $\times 10^8$ &  0.01820 $\times 10^8$\\
2 & $I_3 \rightarrow I_2$ &  9.62296 $\times 10^8$ &  0.01725 $\times 10^8$\\
3 & $I_3$                 &  9.62269 $\times 10^8$ &  0.01698 $\times 10^8$\\
4 & $I_3 \rightarrow I_2$ &  9.62595 $\times 10^8$ &  0.02025 $\times 10^8$   \\
5 & $I_3 \rightarrow I_2$ &  9.62479 $\times 10^8$ &  0.01909 $\times 10^8$     \\
6 & $I_3 \rightarrow I_2$ &  9.62420 $\times 10^8$ &  0.01849 $\times 10^8$   \\
7 & $I_3 \rightarrow I_2$ &  9.62350 $\times 10^8$ &  0.01780 $\times 10^8$\\
8 & $I_3 \rightarrow I_4$ &  9.62494 $\times 10^8$ &  0.01924 $\times 10^8$    \\
9 & $I_3 \rightarrow I_2$ &  9.62526 $\times 10^8$ &  0.01956 $\times 10^8$    \\
10& $I_3 \rightarrow I_4$ &  9.62608 $\times 10^8$ &  0.02037 $\times 10^8$   \\
\hline\hline
\end{tabular}
  \caption{The energy of interstitial defects created at different
  initial positions within the spherical crystal. The energies shown
  are for a power law potential with $s=6$} \label{EI}
\end{table}

It is also informative to track the position dependence of the
interstitial energy after it relaxes. Numerical measurements of
local relaxed energy for the interstitial as a function of initial
location  are presented in Table ~\ref{EI}. We note that, in most
cases, the initial $I_3$ complexion undergoes transitions to more
stable interstitial configurations, except the configuration that
starts from the very center location $C$ in Fig.~\ref{initial}. For
the initial conditions, $C$, $C'$, $0$, and $3$, $I_3$ is the most
stable. We also note that the interstitial created at the exact
center $C$ has the lowest defect energy, while one nearest to the
disclination from the location $10$ requires the largest energy for
interstitial defect formation.

\section{Continuum elastic theory calculations}
\label{5}

In this section we study interstitials analytically using continuum
elastic theory on the sphere. Depending on where we create
interstitials within a spherical crystal, we have seen that they
lead to different annealed states. One can also understand such
position dependency of interstitial defects by calculating the
elastic interaction energy of, for example, the threefold symmetric
interstitial $I_3$ with the embedded 12 disclinations. Here, we
model an interstitial by a compact 5-7-5-7-5-7 ``ring" structure
before the fractionization of the bound dislocations into three
isolated 5-7 dislocations. If ${\bf x}$ represents a coordinate
system within the sphere, the stress field associated with such
configuration is determined by the bi-harmonic equation of a sphere
in terms of the Airy stress function $\chi({\bf x})$, whose sources
are the defect density of disclinations $S_D({\bf x})$, the defect
density of interstitials $S_I({\bf x}) $, and the Gaussian curvature
$K({\bf x})$ of the background crystal, \be \frac{1}{Y}\Delta^2 \chi
({\bf x}) = S_D ({\bf x}) + S_I ({\bf x})-K({\bf x}) \ , \ee where
$Y$ is the two-dimensional Young's modulus \cite{BNT}.  The defect
densities of disclinations and interstitials on a curved crystalline
background are, respectively, written as \be \label{SD} S_D ({\bf
x}) =\frac{\pi}{3 \sqrt{g({\bf x})}} \sum ^{N}_{\alpha =1}
q_{\alpha} \delta({\bf x}- {\bf x}_{\alpha}) \ , \ee and \be
\label{SI} S_I ({\bf x}) = \frac {1}{2\sqrt{g({\bf x})}}
\sum^{N}_{\beta=1} \Omega_{\beta} \Delta \delta({\bf x}- {\bf
x}_{\beta})  \ , \ee where $g({\bf x})$ is the determinant of the
metric tensor of the sphere, $q_{\alpha}$ is the disclination
charge, and $\Omega_{\beta}$ is the local area change caused by
adding an interstitial~\cite{Nelson}. For 5- and 7-fold
disclinations, we have $q_{\alpha}= +1$ and $q_{\alpha} =-1$,
respectively. The elastic free energy in terms of $\chi({\bf x})$ is
given by \be F=\frac{1}{Y} \int dA (\Delta \chi({\bf x}))^2 \ , \ee
with the area element $dA=d^2{\bf x} \sqrt {g({\bf x})}$. The
interaction energy for a distribution of two different kinds of
defects located at positions $\{{\bf x}_{\alpha}\}$ and $\{{\bf
x}_{\beta}\}$, is then obtained from \cite{BNT} \bea \label{U}
\mathcal{U}({\bf x}_{\alpha},{\bf x}_{\beta}) &=&Y \int dA_{\bf x}
S_D ({\bf x}) \int dA_{\bf y} \frac{1}{\Delta^2_{{\bf xy}}} S_I
({\bf y}) \ . \eea By substituting Eqs.~(\ref{SD}) and (\ref{SI})
into Eq.~(\ref{U}) and integrating by parts, the resulting elastic
interaction of interstitials at position $(\theta_{\beta},
\phi_{\beta})$, with disclinations at position $(\theta_{\alpha},
\phi_{\alpha})$ lying in the spherical crystalline background, then
reduces to \be \label{U} \mathcal{U (\theta_{\alpha},
\phi_{\alpha};\theta_{\beta}, \phi_{\beta})} = \frac{Y \pi }{6}
\sum_{\alpha, \beta }q_{\alpha} \Omega_{\beta} \ \Delta \chi
(\theta_{\alpha}, \phi_{\alpha};\theta_{\beta}, \phi_{\beta}) \ .
\ee Here, the interstitial-disclination interaction potential is
given by \be \Delta \chi (\gamma) = \frac{1}{4 \pi} \left[-
\ln\left(\frac{1-\cos \gamma}{2}\right)-1 \right] \ , \ee with the
angular geodesic distance, $\gamma$, between points
$(\theta_{\alpha}, \phi_{\alpha})$ and $(\theta_{\beta},
\phi_{\beta})$, given by \be \label{angular} \cos \gamma = \cos
\theta_{\alpha} \cos \theta_{\beta} +\sin \theta_{\alpha} \sin
\theta_{\beta} \cos(\phi_{\alpha}-\phi_{\beta}) \ . \ee We note that
this interaction depends logarithmically on distance for small
$\gamma$, and corresponds to a repulsive interaction when both
$q_{\alpha}$ and $\Omega_{\beta}$ are positive, as is the case for a
5-fold disclination interacting with an interstitial.  We expect
that similar results hold in flat space.

\begin{figure}[!h]
\centering
\includegraphics[scale=0.5]{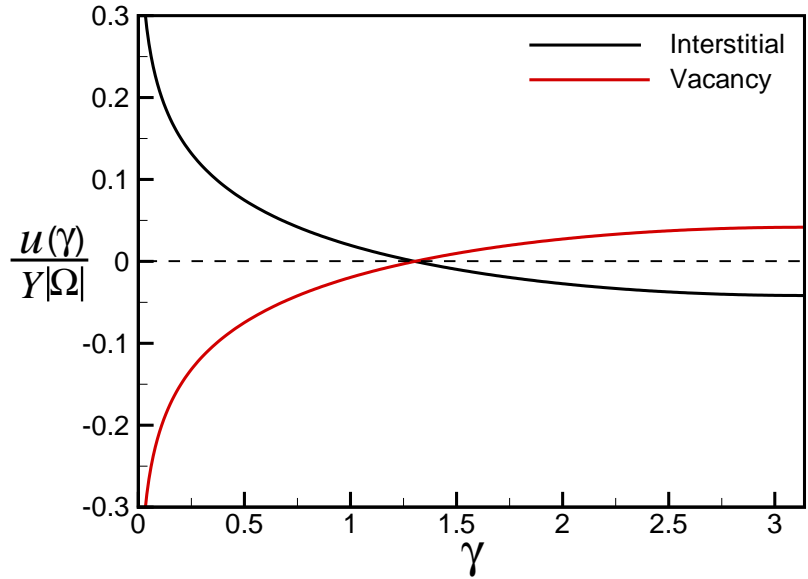}
\caption{The elastic interaction energies of an interstitial and a
vacancy with a single disclination, $\mathcal U (\gamma)$, are
plotted in the unit of $Y|\Omega|$.  } \label{energy}
\end{figure}

To understand the results for a spherical geometry in more detail,
we first plot the contribution to the elastic potential energy
Eq.~(\ref{U}) of a single interstitial as well as a single vacancy
versus the angular distance $\gamma$ to a single 5-fold disclination
with the topological charge $q=+1$ (see Fig~\ref{energy}). We assume
that the magnitudes of the area change $\Omega$ are the same for
both defects. The sign of $\Omega$ can be determined by comparison
with the numerical results. $\Omega
> 0$ corresponds to the interstitial, while $\Omega <0$ to the
vacancy~\cite{Nelson}. The calculation indicates that the elastic
deformation energy associated with the nucleation of an interstitial
falls off with the angular distance from the disclination, whereas
the elastic interaction of a vacancy grows with distance. Thus
interstitials are repelled by a 5-fold disclination (similar to the
discussion of the small $\gamma$ limit above), whereas vacancies are
attracted. Conversely, our calculations suggest that interstitials
are attracted to the cores of $7$-fold disclinations while vacancies
are repelled. This is of interest in constructing ground states on
negatively-curved (hyperbolic) spaces. As noted, however,
interstitials or vacancies typically break up when subjected to
interactions with multiple disclinations.

\begin{figure}
\centering
\includegraphics[scale=0.73]{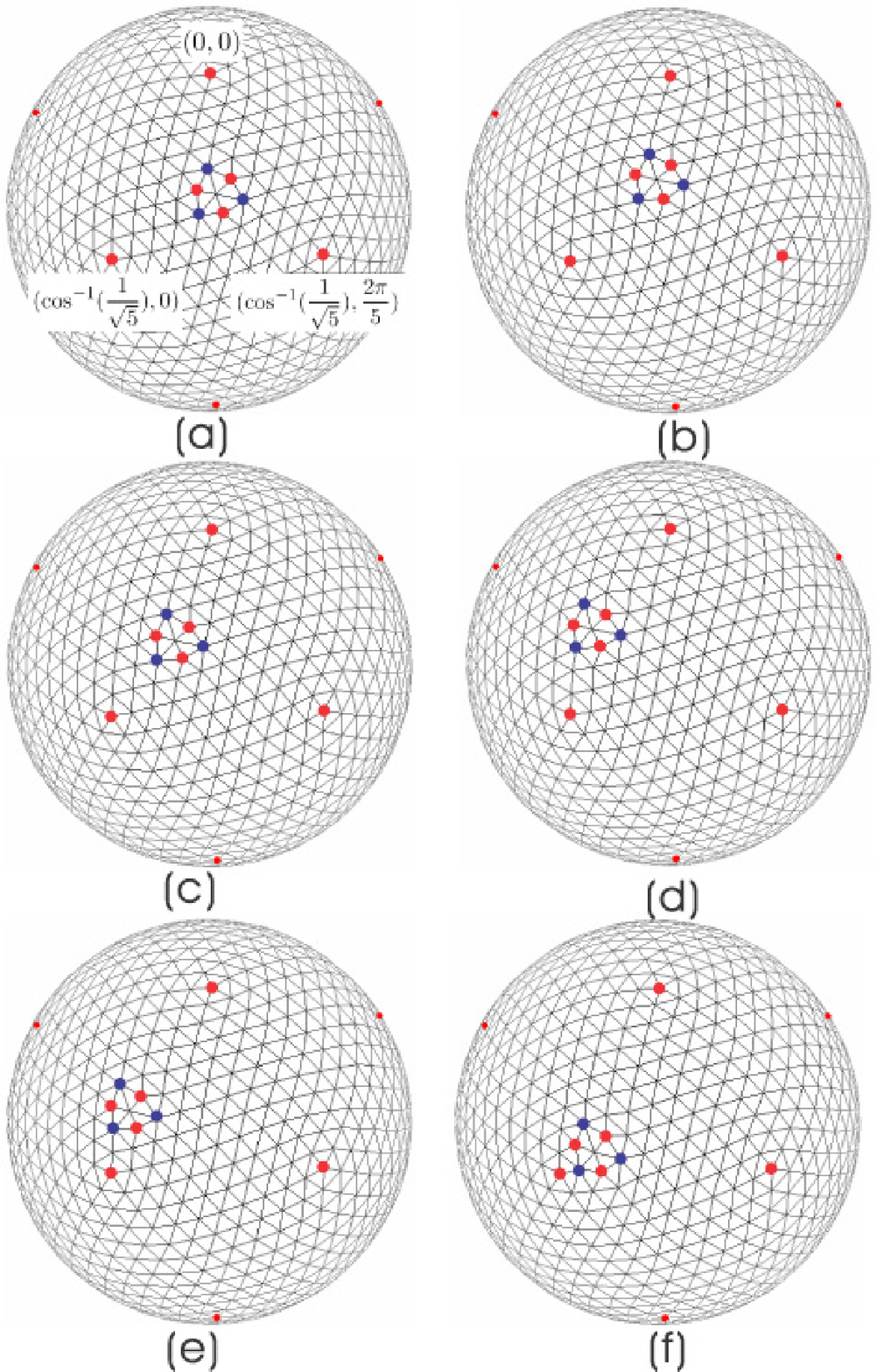}
\caption{Snapshots of interstitials created at various locations
relative to neighboring disclinations. As indicated in part (a), the
disclination at the top vertex is located at the north pole,
$(0,0)$, while those at the bottom left and right are at polar
angles $(\cos^{-1}(1/\sqrt{5}),0)$ and $(\cos^{-1}(1/\sqrt{5},
2\pi/5)$, respectively.  The locations of the center of each
interstitial, $(\theta_I, \phi_I)$, are approximately estimated
using spherical trigonometry: (a) $(0.65, \pi/5)$ (b) $(0.60, 0.54)$
(c) $(0.63,0.33)$ (d) $(0.65,0)$ (e) $(0.77,0)$ (f) $(0.97,0.21)$,
where all angles are in radians. } \label{posii}
\end{figure}

\begin{figure}[!h]
\centering
\includegraphics[scale=0.5]{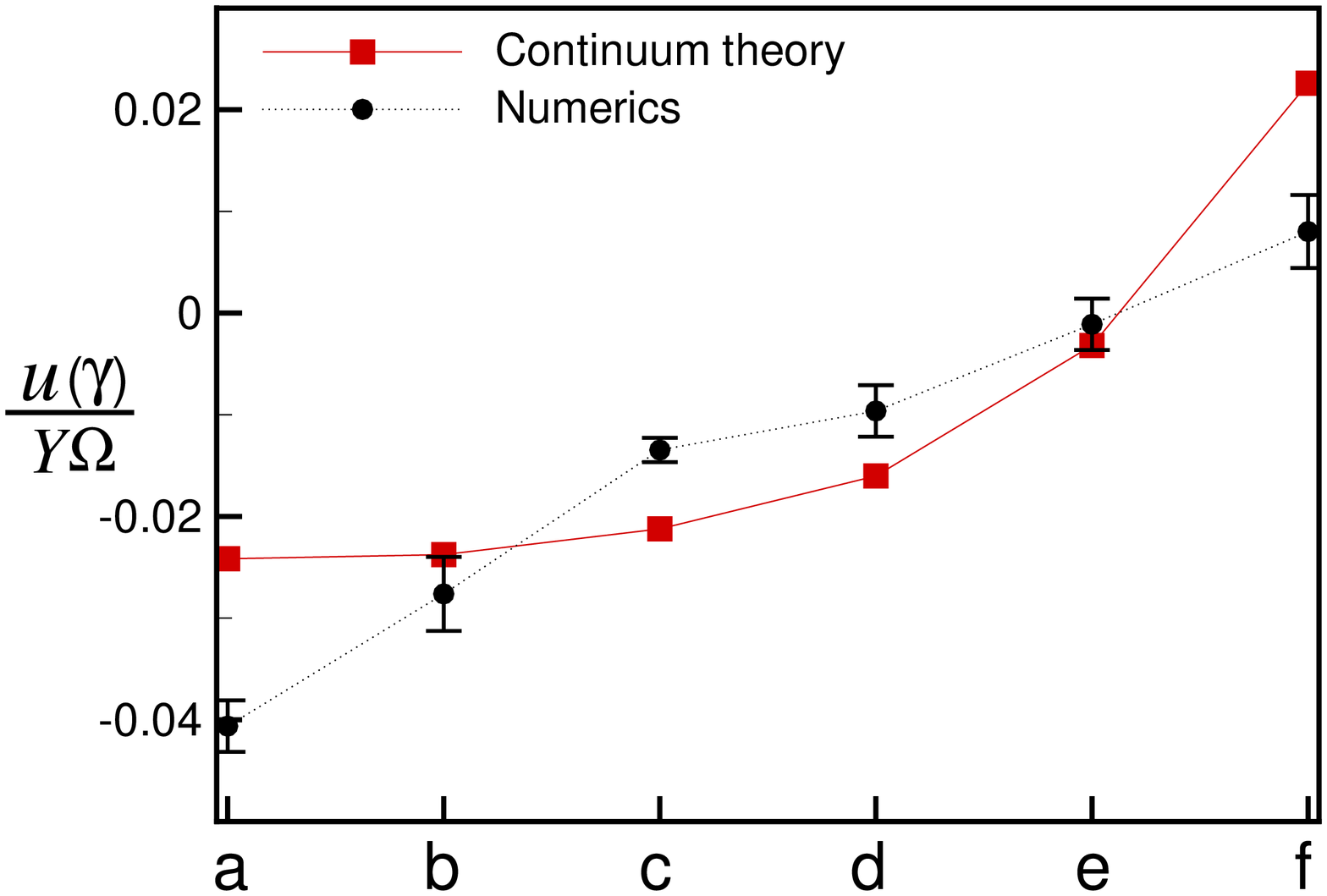}
\caption{For the selected locations of interstitials within the
$(8,3)$ lattice in Fig.~\ref{posii}, the elastic interaction energy
of an interstitial with all twelve disclinations is plotted in the
units of $Y\Omega$, with interstitial locations $a - f$ arranged in
order of increasing $\gamma$. The circles with error bars represent
numerically measured results which are re-scaled to fit the analytic
calculations.} \label{12disc}
\end{figure}

We now calculate the energy of an interstitial interacting with all
twelve disclinations of the spherical crystal in order to compare
this continuum prediction to our numerical results. We first create
interstitials at various locations with snapshots displayed in
Fig.~\ref{posii}. Note that the configurations whose energies are
measured arise from distinct creations of interstitials rather than
from snapshots of the evolution of a single interstitial. In
Fig.~\ref{posii}(a) we show the configuration resulting from adding
a particle at the very center of the spherical triangle (initial
position $C$). The effect of shifting the location of the nucleated
interstitial off the center towards the isolated disclination at the
bottom left is shown in Figs.~\ref{posii}(b) through (e). Finally
Fig.~\ref{posii}(f) illustrates the case of creation immediately
adjacent to the disclination.  For each nucleation of an
interstitial we obtain the relaxed energy $E_{local}$ after
\emph{local relaxation} immediately prior to unbinding into
individual dislocations (fractionization). The numerical results are
scaled (this amounts to taking the product of the Young's modulus
$Y$ and the extra area $\Omega$ of the interstitial as a fitting
parameter) to compare with the analytic calculation discussed below
and presented, with error bars, in Fig.~\ref{12disc}.

For each configuration in Fig.~\ref{posii}, the continuum elastic
potential energies including all 12 disclinations can be obtained by
inserting the known spherical coordinates of the disclinations. For
the twelve vertices of the icosahedron, we choose the following
explicit coordinates, \be (\theta,\phi) \equiv
\left\{(0,0),\left(\delta, \frac {2\pi k}{5} \right)_{0 \leq k \leq
4}, \left(\pi-\delta, \frac{\pi}{5}+\frac{2\pi k}{5} \right)_{0 \leq
k \leq 4},(\pi,0) \right\} \ , \ee where $\delta =\cos^{-1}(1/\sqrt
{5}) \approx 1.107 $ radian. The first 6 vertices lie in the
northern hemisphere and the remaining 6 in the southern hemisphere.
The location of the center of the interstitial is estimated by a
simple counting of lattice spacings together with spherical
trigonometry: $\cos c = \cos a \cos b + \sin a \sin b \cos C $,
where $a$, $b$, and $c$ are the angular lengths of the sides of the
spherical triangle, and $C$ is the angle of the triangle that faces
the side $c$. The coordinates so obtained are given in the caption
of Fig.~\ref{posii}. The resulting total elastic coupling energies
of the interstitials with all disclinations, corresponding to the
configurations in Fig.~\ref{posii}, are plotted together with the
numerical results in Fig.~\ref{12disc}. As can be seen, there is
reasonable agreement between our numerical results and continuum
elastic theory on the sphere.

\section{Vicinity of the magic number}
\label{6}

Icosadeltahedral configurations of magic numbers $N_{nm}$ of
particles on the sphere are believed to be good approximations to
ground states for relatively small numbers of particles, say $N
\lesssim 300$, interacting via a Coulomb potential~\cite{Altsch}. In
this section we investigate the role of interstitials and vacancies
on the energetics of $N$ particles in the vicinity of the magic
numbers $N_{nm}$.

\begin{figure}[!h]
\centering
\includegraphics[scale=0.4]{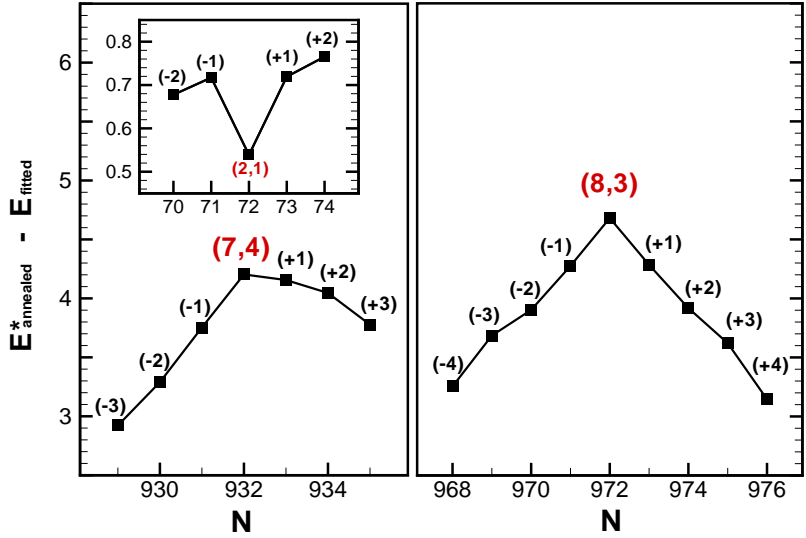}
\caption{We plot the difference between the numerically fitted
energy and the annealed energy with interstitials and vacancies in
the vicinity of the icosadeltahedral configurations with magic
numbers $932$, $972$, and $72$ (inset) for the case of Coulomb
interactions, $s = 1$. The total particle numbers are thus $N_{nm} +
t$, where the integer $t$ is given in parentheses above each data
point. Note that E is dimensionless in our units for this section. }
\label{magic}
\end{figure}

The energy of the system of particles interacting with a power-law
potential Eq.~(\ref{general}) can be expressed as an expansion in
powers of the total number of particles $N$, for $0 < s < 2$, \bea
\label{expansion} 2E_{TOT}(N) = \left[ \frac{N^2}{2^{s-1}(2-s)}- a_1
N^{1 + \frac{s}{2}} + \mathcal{O}(N^{s/2})\right]\frac{e^2}{R^s} \ ,
\eea where the coefficient $a_1$ is a term determined by the
geometry and the microscopic potential~\cite{BCNT2002, BCNT2006}. We
will choose units such that $e^2/R^s$ is unity for our purposes
here. For $s> 2$, the first term in Eq.~(\ref{expansion}), which
arises from long-range interactions, is missing. The coefficient
$a_1$ may be obtained by calculating the correction to the zero mode
energy associated with a particular defect configuration via the
continuum elastic model~\cite{BCNT2006}, or by fitting the energy
from exact minimizations with the function
Eq.~(\ref{expansion})~\cite{Perez}.

To study the total energy in the vicinity of the magic number
$N_{nm}$, it is useful to compare the relaxed energies of particle
numbers $N=N_{nm} \pm t$ along with the inserted interstitial or
vacancy defects $t$, to the expression Eq.~(\ref{expansion}) in
terms of $N$. For this comparison, we use \be
E_{fitted}=\frac{1}{2}(N^2 - 1.10494 N^{3/2}) \ , \ee where $a_1$
was obtained from the icosahedral configuration of twelve 5-fold
disclinations with $s = 1$~\cite{BCNT2006}. Although this value of
$a_1$ is an approximate continuum result, the value above provides a
smoothly varying background energy $E_{fitted}(N)$ useful for our
purposes.

In Fig.~\ref{magic}, we plot $E^*_{annealed} - E_{fitted}$ in the
vicinity of the $(8,3)$ icosadeltahedral lattice with $N_{83}=972$,
$(7,4)$ with $N_{74}=932$, and $(2,1)$ with $N_{21}=72$, where
$E^*_{annealed}$ is the lowest annealed energy we found for
incommensurate particle numbers created initially by inserting
interstitials or vacancies. For large $N$, such as $932$ and $972$,
we find that the icosadeltahedral configuration is a local maximum
rather than a local minimum. Inserting interstitials (or vacancies)
lowers the energy. Presumably, this lowering of the total energy
arises because, as we have shown, vacancies and interstitials
facilitate formation of grain boundary scars, favored for $N \gtrsim
300$. For relatively small numbers of particles, on the other hand,
such as $N=72$, the icosadeltahedral configuration is a local
minimum, consistent with expectations for $N \lesssim
300$~\cite{BNT}. Somewhat similar results near special numbers of
capsids were found in a minimal model for the equilibrium energy of
viral capsids using Monte Carlo simulations~\cite{zandi}.

\vspace{10mm} \centerline{\bf Acknowledgements} The work of MJB and
HS was supported by the NSF through Grants DMR-0219292 and
DMR-0305407, and through funds provided by Syracuse University. The
work of DRN was supported by the Harvard MRSEC through Grant
DMR-0213805.

\end{document}